\documentclass[letterpaper, 10pt,conference]{IEEEtran}
\IEEEoverridecommandlockouts
\usepackage[T1]{fontenc}
\usepackage{cite}
\usepackage{comment}
\usepackage{booktabs}
\usepackage{amsmath,amssymb,amsfonts}
\usepackage{algorithmic}
\usepackage{graphicx}
\usepackage{textcomp}
\usepackage{xcolor}
\usepackage{color,soul}
\usepackage[caption=false,font=footnotesize]{subfig}
\usepackage{upgreek}
\def\BibTeX{{\rm B\kern-.05em{\sc i\kern-.025em b}\kern-.08em T\kern-.1667em\lower.7ex\hbox{E}\kern-.125emX}}

\usepackage{comment}
\usepackage{multirow}
\usepackage{tabularx}
    
\begin{document}

\title{Reinforcement Learning-Based 3D Beam Adaptation for Underwater Wireless Optical Communication with AUVs}

\author{\IEEEauthorblockN{
Viviana Centritto Arrojo,   
Ama Bandara,    
Sergi Abadal,
Evgenii Vinogradov}                                     
\IEEEauthorblockA{Nanonetworking Center in Catalunya, Universitat Polit\`{e}cnica de Catalunya, Barcelona, Spain}

}

\maketitle

\begin{abstract}
Underwater Wireless Optical Communications (UWOC) provide essential high data rates for Autonomous Underwater Vehicles (AUVs), but reliable connectivity is critically affected by transmitter-receiver misalignment. This work addresses the beam pointing problem for a moving AUV subject to unknown ocean currents through a Deep Reinforcement Learning (DRL) framework. We develop a comprehensive 3D UWOC channel model incorporating depth-dependent attenuation, turbulence, and a discrete-ray method to accurately quantify geometric and misalignment losses. The resulting agent jointly optimizes beam steering and divergence angles, learning a policy that prioritizes continuous link maintenance. Evaluated using real oceanographic data, the framework's performance is assessed via the excess outage metric, which isolates outages occurring exclusively due to pointing errors. Relative to perfect alignment between nodes, the proposed method bounds this metric to 8.5\% under ocean current-free tracking conditions and limits it to 16.2\% when subjected to current-induced drift. The findings obtained in this work confirms that DRL-enabled beam control is capable of adaptive tracking and mitigating link outages in dynamic underwater environments.
\end{abstract}



\section{Introduction}

UWOC has emerged as an essential technology for high-capacity underwater links, overcoming the severe attenuation of radio frequency (RF) signals and the bandwidth limitations of acoustic systems~\cite{Gorodilov2025}. Characterized by high data rates and low latency, UWOC is a relevant solution for integrating underwater networks into beyond 5G and 6G ecosystems \cite{saeed2019underwater}.

Nevertheless, the underwater environment presents a channel plagued by impairments, leading to severe problems of pointing accuracy \cite{vali2017use}.While environmental factors degrade signal quality, a lack of alignment between the transmitter and receiver can produce a complete link outage. This alignment is further compromised by ocean currents, which introduce drift and jitter, resulting in significant positional uncertainty \cite{saeed2019underwater}, \cite{shakeera2023ml}. In this context, there is a need for autonomous systems capable of processing high amounts of heterogeneous information to make optimal real-time decisions. The objective of such systems is to ensure the reliability of the communication link in harsh dynamic environments. 

Recent literature has demonstrated the efficacy of reinforcement learning (RL) algorithms in addressing UWOC pointing challenges. Romdhane and Kaddoum \cite{romdhane2022reinforcement} proposed a dynamic RL beam width orientation and adaptation framework. This was achieved by implementing model-free Q-learning and State-Action-Reward-State-Action (SARSA) algorithms to optimize link performance for underwater nodes. To improve upon reactive approaches, the work in \cite{ishida2025adaptive} integrates Long Short-Term Memory (LSTM) with Deep-Q Networks 
(DQNs), leveraging the LSTM to predict vessel displacement based on historical wave data, and enabling the DQN agent to optimize both the beam divergence and steering angles. Furthermore, to maintain a seabed-to-vessel link disrupted by waves, the authors in \cite{shin2024multidimensional} propose a two-agent DRL framework. The system divides the tracking tasks temporally: one agent updates the beam's azimuth and elevation over longer intervals, while a second agent preemptively fine-tunes the divergence at every time slot to ensure coverage.

However, prior studies largely rely on 2D spatial models or focus on surface-bound receivers disrupted by surface waves. By treating the underwater channel as a homogeneous environment, they cannot model depth-dependent variations in attenuation, turbulence and noise. Consequently, 3D tracking for submerged, mobile AUVs remains unresolved. We propose a dynamic 3D DRL framework that incorporates vertical link characteristics and underwater current-induced positional uncertainty. To accurately quantify geometric and pointing losses, our approach evaluates the individual optical rays failing to reach the receiver aperture.

Therefore, the main contributions of this work are: \textbf {(i)}~characterization of the 3D UWOC environment; and \textbf{(ii)} design and evaluation of a RL-based DQN control framework that maintains beam alignment and link quality.



\section{System Model} 
\label{sec:system}

\begin{figure}[!t] 
    \centering    
    \subfloat[]{
        \includegraphics[width=0.50\columnwidth]{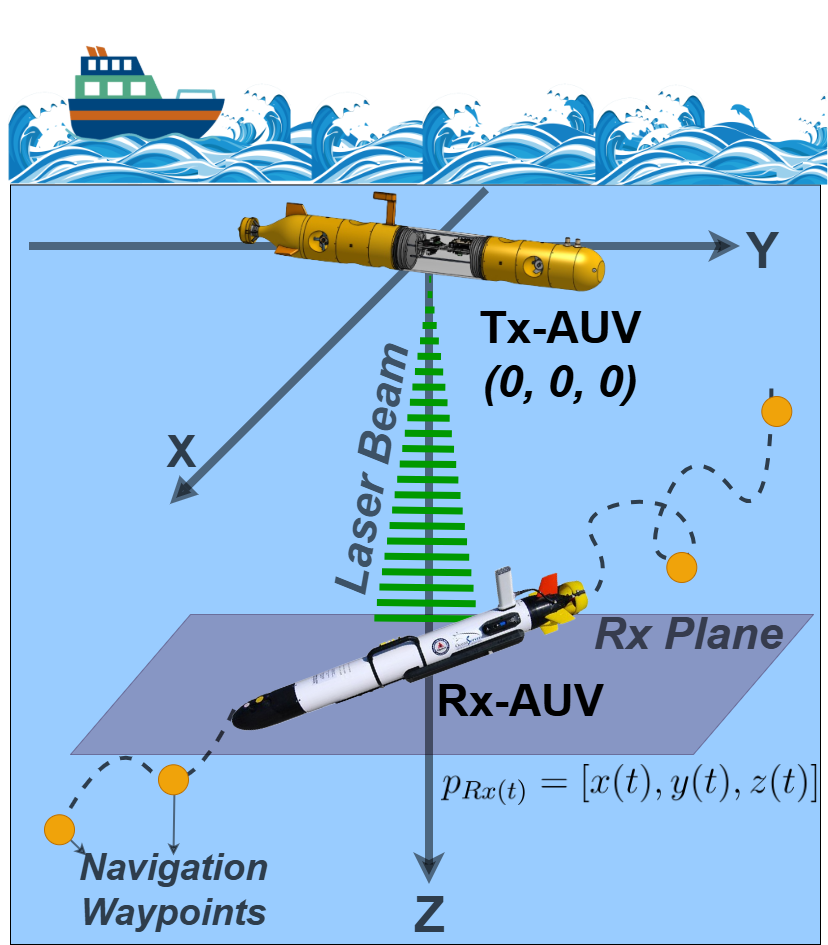}
        \label{fig:uwocsystem}
    }\hfill 
    \subfloat[]{
        \includegraphics[width=0.35\columnwidth]{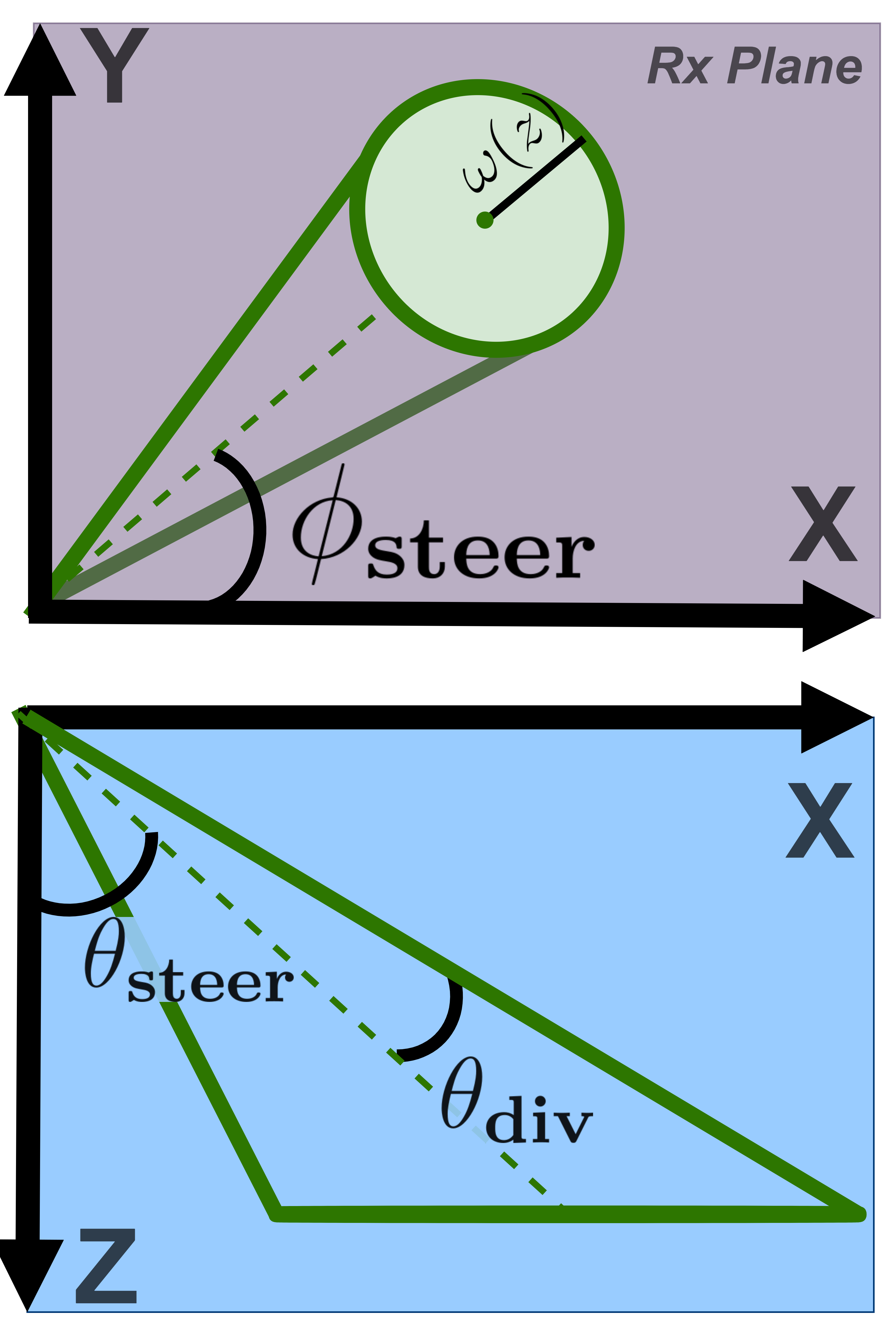}
        \label{fig:uwoc_angle_model} 
    }    
    \caption{UWOC system model. (a) 3D downlink setup between the fixed T$_x$ and mobile R$_x$ AUVs. (b) Geometric representation of the controllable beam steering and divergence angles.}
    \label{fig:system_model}
    \vspace{-0.4cm}
\end{figure}

Fig. \ref{fig:uwocsystem} illustrates the point-to-point downlink UWOC communication scenario between a stationary transmitter (T$_x$) AUV near the surface and a mobile receiver (R$_x$) AUV. We adopt a 3D Cartesian coordinate system with depth measured positively along the downward $z$ axis. The R$_x$-AUV follows a dynamic trajectory with an instantaneous position \mbox{$p_{\text{Rx}}(t) = [x(t), y(t), z(t)]$}, defined using a set of $N$ discrete waypoints $W_k = [t_k, x_k, y_k, z_k]$ for $k = 1, \dots, N$. Furthermore, the T$_x$-AUV emits a Gaussian laser beam with wavelength $\lambda$, beam radius $w_z$, and half-divergence angle $\theta_{\text{div}}$. As shown in Fig.~\ref{fig:uwoc_angle_model}, the T$_x$ steers the beam axis toward any location using polar $\theta_{\text{steer}}$ and azimuthal $\phi_{\text{steer}}$ angles. To maintain the communication link, the T$_x$ tracks the R$_x$-AUV's trajectory by adjusting the beam steering and divergence angles. However, because underwater currents perturb the R$_x$ from its original trajectory, we represent these disturbances as depth-dependent velocity components, namely a northward velocity $V_v$ ($y$-direction) and an eastward velocity $V_h$ ($x$-direction) \cite{shakeera2023ml}.

\begin{figure}[!t] 
    \centering   
    \includegraphics[width=0.95\linewidth]{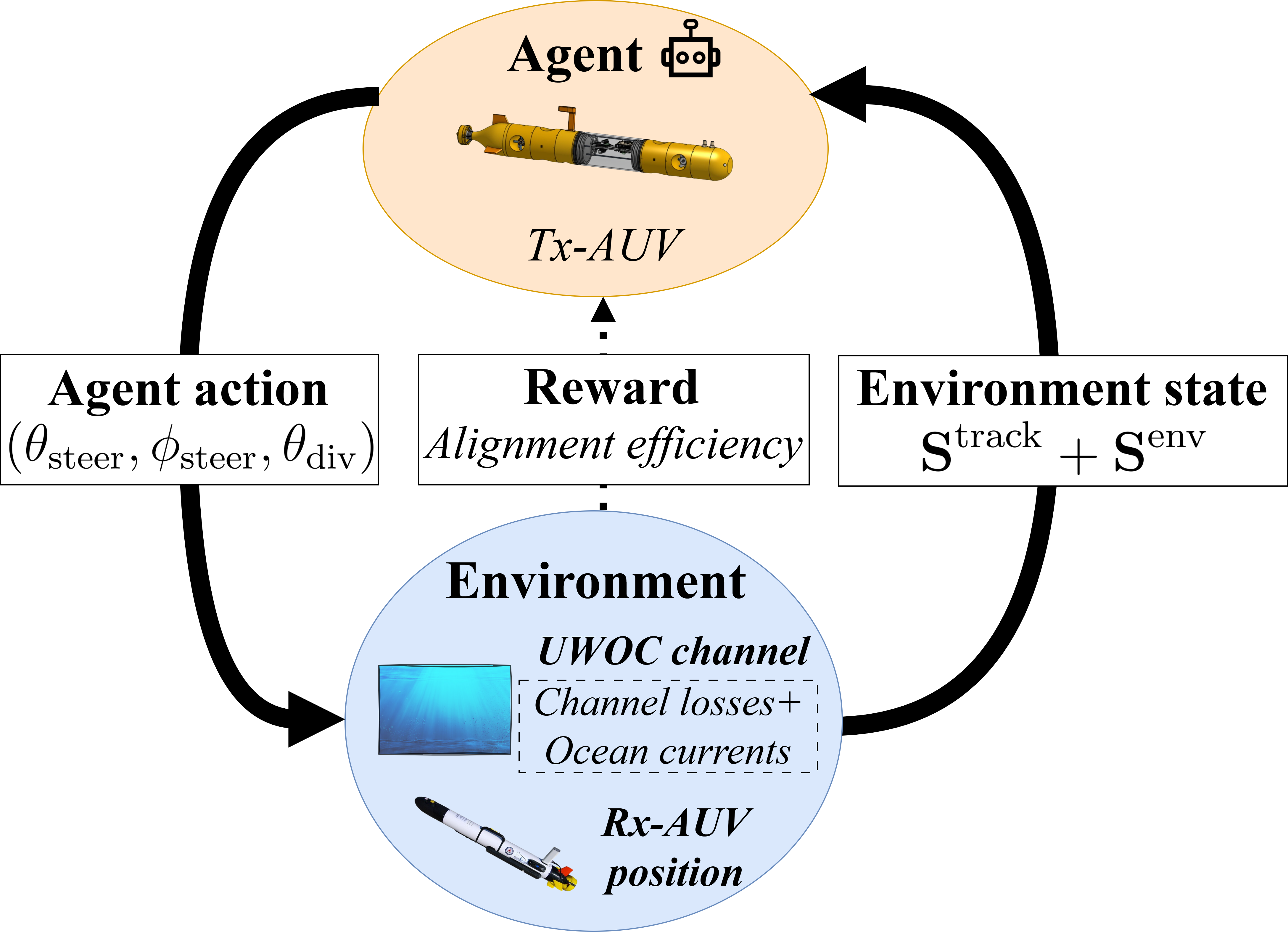}
    \caption{DRL framework illustrating the closed-loop interaction between the agent and UWOC environment. The T$_x$-AUV agent optimizes the beam steering and divergence angles based on the tracking and environmental states.}
    \label{fig:DRL_framework}
\end{figure}

The DRL framework is depicted in Fig. \ref{fig:DRL_framework}, where a RL agent at the T$_x$-AUV governs the steering and divergence mechanisms. Based on state observations, the agent selects actions that interact with the UWOC environment. During training, the environment evaluates these actions and returns a reward signal reflecting the agent's efficiency. The following assumptions are considered: \textbf{(i)} both AUVs are initialized with knowledge of the starting positions and planned waypoints; \textbf{(ii)} each vehicle is equipped with an Inertial Measurement Unit (IMU) for onboard position and velocity estimation; \textbf{(iii)} since the AUVs can operate collaboratively, they exchange state information via a bidirectional UWOC link. While this work focuses on the T$_x$-side agent, the same procedure can be performed at the R$_x$-AUV when the communication roles are reversed.

To validate the proposed framework, we evaluate our model under two distinct operational scenarios. The first is a current-free environment that assumes zero positional uncertainty. The second scenario introduces a dynamic drift caused by ocean currents, forcing the agent to continuously update the R$_x$'s position estimates based on historical and environmental data.

\section{DRL implementation} 
\label{sec:RL_method}
The DRL model maintains a consistent action space, reward function, and UWOC environment across both scenarios, which differ solely in their state space formulation.
\subsection{State space vector (\textbf{$S_t$}) } 
Under the current-free scenario, $S_t$ is represented as $S^{\text{track}}$, which encompasses 11 variables tailored to address the R$_x$ tracking problem. However, to model the drift caused by ocean currents, we leverage the Hybrid Coordinate Ocean Model (HYCOM) \cite{nrl_gofs31} to generate a dataset of the dynamics influencing the R$_x$-AUV's trajectory. This requires expanding the agent's observation space to incorporate the environmental data. The augmented $S_t$ consists of 24 continuous variables divided into two branches, $S^{track}_t$ and $S^{env}_t$, introducing 13 new components. This decoupling separates the beam dynamics from environmental uncertainties, as detailed in Table \ref{tab:state-space}.
\begin{table}[!t]
    \centering
    \caption{Features defining state space vectors}
    \label{tab:state-space}
    \renewcommand{\arraystretch}{1.3} 
    \begin{tabularx}{\columnwidth}{|c|c|X|}
        \hline
        \textbf{Vector} & \textbf{Component} & \textbf{Description} \\
        \hline
        \multirow{3}{*}{$S^{\text{track}}$} 
        & $\mathbf{p}_t, \mathbf{p}_{t-1}$ & Receiver position at time $t$ and $t-1$ in spherical angles $(\theta,\phi,r)$  \\ \cline{2-3}
        & $P_{Rx,t}, P_{Rx,t-1}$ & Received optical power at time $t$  and the previous time step. \\ \cline{2-3}
        & $\mathbf{b}_t$ & Beam parameters at time $t$: Agent's beam parameter settings ($\theta_{steer}, \phi_{steer}, \theta_{div}$). \\
        \hline
        \multicolumn{2}{|c|}{\textbf{Total Dimension}} & \textbf{11 components} \\
        \hline
        \hline 
        \multirow{7}{*}{$S^{\text{env}}$} 
        & $z_t$ & Depth of the receiver at time step $t$. \\ \cline{2-3}
        & $\eta_{surf}$ & Sea surface elevation. \\ \cline{2-3}
        & $\mathbf{d}_{cyc}$ & Cyclical day encoding vector. \\ \cline{2-3}
        & $\mathbf{h}_{cyc}$ & Cyclical hour encoding vector. \\ \cline{2-3}
        & $\mathbf{v}_{lag3}, \mathbf{v}_{lag24}$ & Water x-y velocity from 3 and 24 hours prior. \\ \cline{2-3}
        & $\mathbf{w}_{lag3}$ & Water temperature and salinity from 3 hours prior. \\ \cline{2-3}
        & $\tau_t$ & Time step $t$ of the episode. \\
        \hline
        \multicolumn{2}{|c|}{\textbf{Total Dimension}} & \textbf{13 components} \\
        \hline
    \end{tabularx}
\end{table}
\vspace{-0.45cm}
\subsection{Action space}
The agent operates within a discrete action space comprising 27 distinct actions representing every possible combination of increasing, decreasing, or maintaining the beam controllable parameters: the steering angles ($\theta_{\text{steer}}$ and \, $\phi_{steer}$) for the pointing direction, and the divergence angle $\theta_{\text{div}}$ for the beam width. At each time step, the agent updates each parameter via an action selected from $[up, keep, \, down]$ with step sizes \mbox{$\Delta_{\theta} = \Delta_{\phi}$ = 10 mrad}, and $\Delta_{\text{div}}$ = 1 mrad. To compute Q-values for these 27 actions, the current-free scenario uses a standard Multilayer Perceptron (MLP), while the induced-current case employs a two-tower MLP to independently process $S^{track}$ and $S^{env}$.
\subsection{Reward function}
\begin{figure}[!t] 
    \centering   
    \includegraphics[width=1\linewidth]{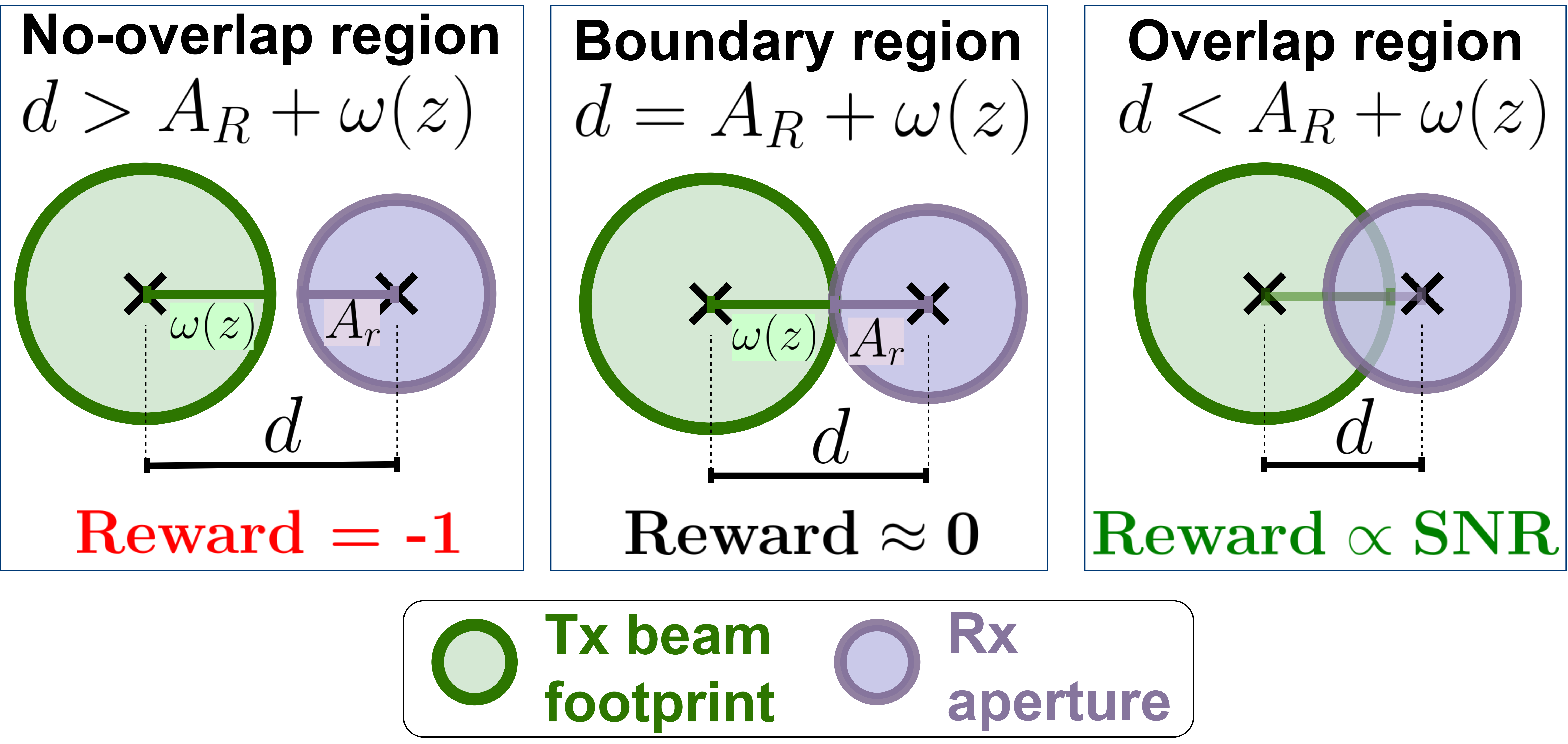}
    \caption{Geometric constraint of $R_t$ based on the spatial overlap between the beam radius $w(z)$ and the R$_x$ aperture radius $A_r$.}
    \label{fig:reward_condition}
    \vspace{-0.3cm}    
\end{figure}

To maintain a robust link while penalizing pointing errors, we define the reward function as
\begin{equation}
    R_t = 
    \begin{cases} 
        \frac{SNR_{t}}{SNR_{ideal}} & \text{if } d \le [A_{r} + w(z)] \\
        -1 & \text{if } d > [A_{r} + w(z)],
    \end{cases}
\end{equation}
where $d$ denotes the positional misalignment between the optical beam center and the R$_x$-AUV aperture center, and $A_{r}$ is the R$_x$ aperture radius. This approach prioritizes alignment efficiency, defined as the ratio of the instantaneous Signal-to-Noise Ratio $SNR_{t}$ to the maximum possible $SNR_{ideal}$ at a given depth $z$, eliminating the depth-dependence signal attenuation inherent to the underwater channel. 
The reward $R_t$ is subject to a geometrical constraint, as illustrated in Fig. \ref{fig:reward_condition}. If $d$ exceeds the sum of $w_{z}$ and $A_{r}$, the R$_x$-AUV fails to capture any optical power. To discourage this outcome, the agent receives a penalty of $-1$. Notably, while $A_{r}$ represents a fixed hardware constraint, $w(z)$ is controlled by the agent. 

\section{UWOC Environment: Channel Modeling} 
\label{sec:uwoc_modeling}
The electrical signal received at the photodetector is modeled as \mbox{$r = \eta h s + n$}, where $s$ denotes the transmitted optical power, $\eta$ is the photodetector responsivity, $h$ is the overall channel loss coefficient, and $n$ represents the additive noise components. This loss is decomposed as $h = h_{\text{c}} h_{\text{t}} h_{\text{g}} h_{\text{p}}$, which respectively account for channel attenuation $h_{\text{c}}$, turbulence $h_{\text{t}}$, geometric spreading $h_{\text{g}}$, and pointing losses $h_{\text{p}}$ \cite{cox2014simulating}.

\subsection{Channel attenuation}
The attenuation model is based on Koutsonas \textit{et al.} \cite{koutsonas2026performance}. It follows Beer-Lambert Law (BLL), which states that power decays exponentially with respect to the propagation distance $L_{t}$ and attenuation coefficient $c(\lambda, z)$ \cite{saeed2019underwater}. Depending on water type and depth $z$, this coefficient is the sum of the absorption $a(\lambda, z)$ and scattering coefficients $b(\lambda, z)$ \cite{jerlov1976marine}. Both $a(\lambda, z)$ and $b(\lambda, z)$ are based on the Haltrin model \cite{Haltrin}. The absorption coefficient is modeled by accounting for the individual contributions of several aquatic constituents, each scaled by its concentration $C$ \cite[Eq. (1)]{Haltrin}
\begin{equation}
a(\lambda, z) = {} C_{w}a_w(\lambda,z) + C_{phy}a_{phy}+ C_{g}a_g(\lambda,z) + C_{n}a_n(\lambda,z),
\end{equation}
where the subscripts $w, phy, g, \text{and } n$ denote pure water, chlorophyll-dominated phytoplankton \cite{Deng}, colored dissolved organic material, and non-algal suspended particles \cite{johnson2015optical}, respectively. In this work, only pure water and chlorophyll are considered as the main absorption components. Furthermore, the scattering coefficient is calculated as the summation of scattering contributions from pure sea water $b_w$, small particles $b_s$ and large organic particles $b_l$ \cite[Eq. (2)]{Haltrin}
\begin{equation}
b(\lambda, z) = b_w + C_{s}(z)b_s(\lambda) + C_{l}(z)b_l(\lambda),
\end{equation}
where $C_s(z)$ and $C_l(z)$ represent the effective concentrations of these particles as estimated by \cite[Eq. (14)]{Haltrin}
\begin{align}
C_s(z) &= 0.01739 \cdot C_{phy}(z) \cdot \exp\left[ 0.11631 \cdot C_{phy}(z) \right] \\
C_l(z) &= 0.76284 \cdot C_{phy}(z) \cdot \exp\left[ 0.03092 \cdot C_{phy}(z) \right].
\end{align}

The chlorophyll concentration $C_{phy}(z)$ is linked to $a(\lambda,z)$ and $b(\lambda,z)$. To capture the vertical heterogeneity of the water column induced by the Deep Chlorophyll Maximum (DCM), $C_{phy}(z)$ is modeled using a Gaussian vertical profile. This profile is parameterized by the surface background chlorophyll concentration $B_0$ and its vertical gradient $S$, the chlorophyll peak depth $z_{max}$, the integrated peak biomass $h$, and the DCM layer thickness $\sigma$ \cite[Eq. (7)]{Utiz}
\begin{equation}
C_{phy}(z) = B_0 + S \cdot z + \frac{h}{\sigma \sqrt{2\pi}} \exp\left[ -\frac{(z - z_{max})^2}{2\sigma^2} \right].
\end{equation}
By adjusting $B_0$, the model simulates four distinct water types: pure sea water ($B_0 =$ 0.005), clear ocean ($B_0 =$ 0.31), coastal ($B_0 =$ 0.83) and harbor ($B_0 =$ 5.9) \cite{Utiz,koutsonas2026performance}. 

\subsection{Turbulence}
As shown in Fig. \ref{fig:turbulence}, the turbulence model divides the water column into $K$ discrete layers with each layer characterized by its thickness $d_k$, and scintillation index $\sigma_{h_{\text{t},k}}^2$. Under weak turbulence conditions, the turbulence loss is modeled as a cascaded fading channel where the total fading coefficient $h_{\text{t}}$ is the product of the independent layer coefficients $h_{\text{t},k}$ \cite[Eq.~(1)]{elamassie2020vertical}
\begin{equation}
h_{\text{t}} = \prod_{k=1}^{K} h_{\text{t},k}.
\end{equation} 
Since each $h_{\text{t},k}$ is a lognormal random variable, $h_{\text{t}}$ is inherently lognormal. This parameter is determined using a multi-step procedure based on \cite{elamassie2020vertical}. 

\subsubsection{Characterization of the turbulence environment} Temperature and salinity profiles are extracted from HYCOM~\cite{nrl_gofs31}. Using these profiles, the density ratio $\omega$ is computed as~\cite[Eq. (19)]{elamassie2020vertical}
\begin{equation}
\omega = \frac{\alpha (dT/dz)}{\beta (dS/dz)},
\end{equation}
where $dT/dz$ and $dS/dz$ are the vertical gradients of temperature and salinity, respectively. The parameter $\alpha$ represents the thermal expansion coefficient and $\beta$ is the saline contraction coefficient of each layer; both are calculated using the Gibbs-SeaWater (GSW) Oceanographic Toolbox \cite{TEOS10_Software}. 

\subsubsection{Calculation of the scintillation index $\sigma_{h_{\text{t},k}}^2$} Each layer's scintillation index is derived from its Rytov variance $\sigma_{1_{k}}^{2}(d_k)$ 
\cite[Eq. (19)]{tayebnaimi2024scintillation}
\begin{equation}
\sigma_{ht,k}^{2} = \exp(4\sigma_{1_{k}}^{2}(d_k)) - 1,
\end{equation}
which requires numerical integration over the normalized path distance $L$ and spatial frequency $\kappa$ \cite[Eq. (20)]{tayebnaimi2024scintillation}
\begin{equation}
\begin{split}
\sigma_{1_k}^2(d_k) = {} & 2\pi^2 k_1^2 d_k \int_0^1 \int_0^\infty \kappa \Phi_{n_k}(\kappa) \exp\left(\frac{-\kappa^2 \rho^2 d_k \Lambda_{1, k}}{k_1}\right) \\
& \times \left(1 - \cos \left( \frac{\kappa^2 d_k \rho(1-\rho) \hat{\Theta}_{1, k}}{k_1} \right) \right) \,\mathrm{d}\kappa \,\mathrm{d}\rho,
\end{split}
\end{equation} 
where the key auxiliary parameters include the optical wavenumber $k_1$, the modified Nikishov spatial power spectrum for a Gaussian beam through the $k$-th layer $\Phi_n(\kappa)$, and both the Fresnel ratio $\Lambda_{1, k}$ and estimated curvature parameter $\hat{\Theta}_{1, k}$ evaluated at that layer's upper boundary.

\subsubsection{Estimation of the total turbulence fading coefficient} Assuming statistically independent fading across the layers, we determine the total scintillation index by summing the individual log-variances of each discrete layer                   \cite[Eq. (3)]{elamassie2020vertical}
\begin{equation}
\sigma_{h_t}^2 = \sum_{k=1}^{N} \ln(1 + \sigma_{h_{t,k}}^2).
\end{equation}
Finally, we stochastically generate $h_{\text{t}}$ as 
\begin{equation}
h_{\text{t}} = \exp\left(\mu_{\text{t}} + \mathcal{N}(0, 1) \cdot \sigma_{h_{\text{t}}}\right),
\end{equation} 
where the mean logarithmic amplitude $\mu_t$ is normalized to ensure a  unity power expectation \cite{elamassie2020vertical}.
\subsection{Geometric spreading and pointing losses}
\begin{figure}[!t]
    \centering   
    \subfloat[]{
        \includegraphics[width=0.4\columnwidth]{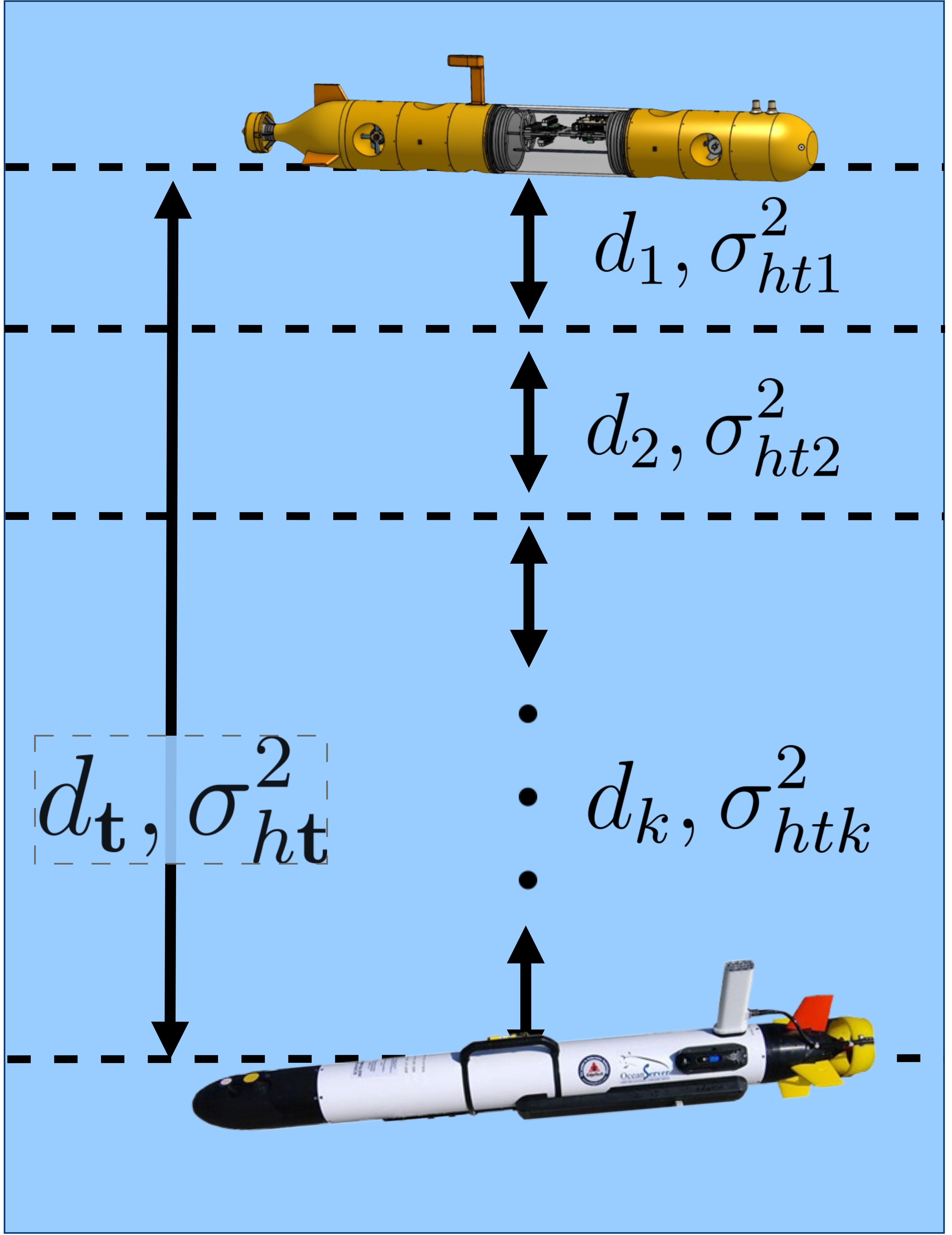}
        \label{fig:turbulence}
    }       
    \hfill
    \subfloat[]{
        \includegraphics[width=0.50\columnwidth]{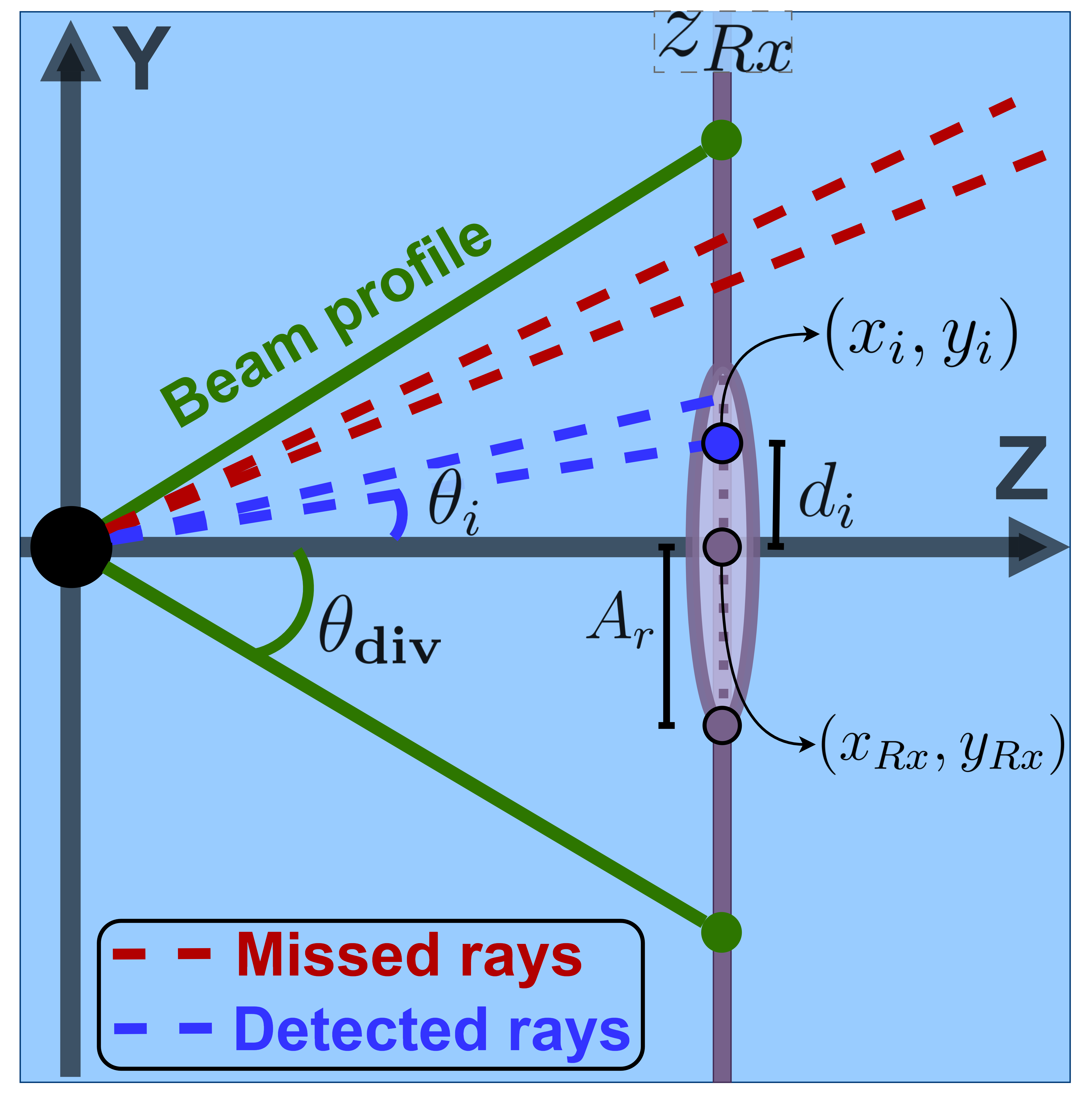}
        \label{fig:ray_misalignment}
    }    
    \caption{UWOC channel impairments. (a) Cascaded turbulence model with depth-dependent layers ($d_{k}$, $\sigma_{h_{\text{t},k}}^2$). (b) Discrete-ray method to calculate geometric spreading and pointing losses.}
    \label{fig:environment_models}
\end{figure}

The laser beam profile is modeled using $N$ discrete rays. Fig. \ref{fig:ray_misalignment} illustrates the initial ray trajectories, defined by spherical angles $(\theta_i, \phi_i)$ to reproduce a Gaussian intensity profile with a divergence angle $\theta_{\text{div}}$ \cite{cox2014simulating}. These rays are then projected onto the R$_x$ plane at depth $z_{\text{R}x}$, yielding intersection points $P_{\text{int}}^{(i)} = (x_i, y_i, z_{\text{R}_x})$. Geometric loss arises from the spatial spreading of the rays over the propagation distance. For a ray to be detected, the distance $d_i$ of the $i$-th ray's intersection point $P_{\text{int}}^{(i)}$ to the center of the R$x$-AUV aperture $(x_{Rx}, y_{Rx})$ must be bounded by the receiver radius $A_r$ 
\begin{equation}
d_i = \sqrt{(x_i - x_{Rx})^2 + (y_i - y_{Rx})^2} \leq A_r.
\end{equation}
The proportion of rays failing to meet this condition quantifies the combined effect of geometric and pointing losses.

\subsection{Noise model}
To account for the primary noise components specific to the oceanic environment, we define the total noise variance $\sigma_\text{total}^2$ as the sum of three independent variances: ambient solar background $\sigma_\text{solar}^2$, dark current $\sigma_\text{dark}^2$, and thermal $\sigma_\text{thermal}^2$ \cite{zayed2025modeling,mahmutouglu2020investigation}. Among these, $\sigma_\text{solar}^2$ often constitutes the most significant source of interference~\cite{johnson2015optical}. This variance is calculated as a function of the electronic charge $q$, the R$_x$ sensitivity $\mathcal{S}$, the bandwidth $B$ and the background solar power $P{\text{sol}}$ \cite[Eq. (4), Eq (5)]{mahmutouglu2020investigation}
\begin{equation}
\sigma_{\text{solar}}^2 = 2 q \mathcal{S} P_{\text{sol}} B,
\end{equation}
where $P_{\text{sol}}$ depends on the R$_x$'s field of view (FOV), aperture radius, and the upwelling solar radiance at the R$_x$'s depth.

\section{Results}
\label{sec:results}
The assessment of the proposed framework begins with a tailored training phase, utilizing 15,000 episodes for the current-free scenario and 30,0000 episodes for the current-induced case. Each 10-s episode comprises 100 discrete steps divided into 0.1-s decision intervals. At the start of every episode, a unique R$_x$ trajectory is generated via an interpolated 3D random walk, bounded between 1 and 50~m in depth with simulated AUV velocities ranging from 0.6 to 1.2~m/s. The scenario specific DRL hyperparameters and the share UWOC simulation parameters are summarized in Table \ref{tab:training_hyperparameters}. 

\begin{table}[!t]
\centering
\renewcommand{\arraystretch}{1.3} 
\caption{UWOC simulation parameters and DRL hyperparameters}
\label{tab:training_hyperparameters}
\begin{tabular}{|c|l|c|c|}
\hline
\textbf{Category} & \multicolumn{1}{c|}{\textbf{Parameter}} & \begin{tabular}{@{}c@{}}\textbf{Current} \\ \textbf{free}\end{tabular} & \begin{tabular}{@{}c@{}}\textbf{Current} \\ \textbf{induced}\end{tabular} \\
\hline
\multirow{5}{*}{DRL} 
& Learning rate & $10^{-4}$ & $5 \times 10^{-5}$ \\ \cline{2-4}
& Batch size & $64$ & $128$ \\ \cline{2-4}
& Update frequency & $100$ & $200$ \\ \cline{2-4}
& Replay buffer & $50,000$ & $100,000$ \\ \cline{2-4}
& $\epsilon$-decay & $10^{-4}$ & $3 \times 10^{-6}$ \\
\hline
\hline
\multirow{3}{*}{UWOC} 
& Chlorophyll $B_0$ & \multicolumn{2}{c|}{$0.005$} \\ \cline{2-4}
& Wavelength $\lambda$ & \multicolumn{2}{c|}{$530$~nm} \\ \cline{2-4}
& $R_x$ aperture $A_r$ & \multicolumn{2}{c|}{$0.2$~m} \\
\hline
\end{tabular}
\end{table}

\begin{figure} [!t]
\centering
\includegraphics[width=1\columnwidth]{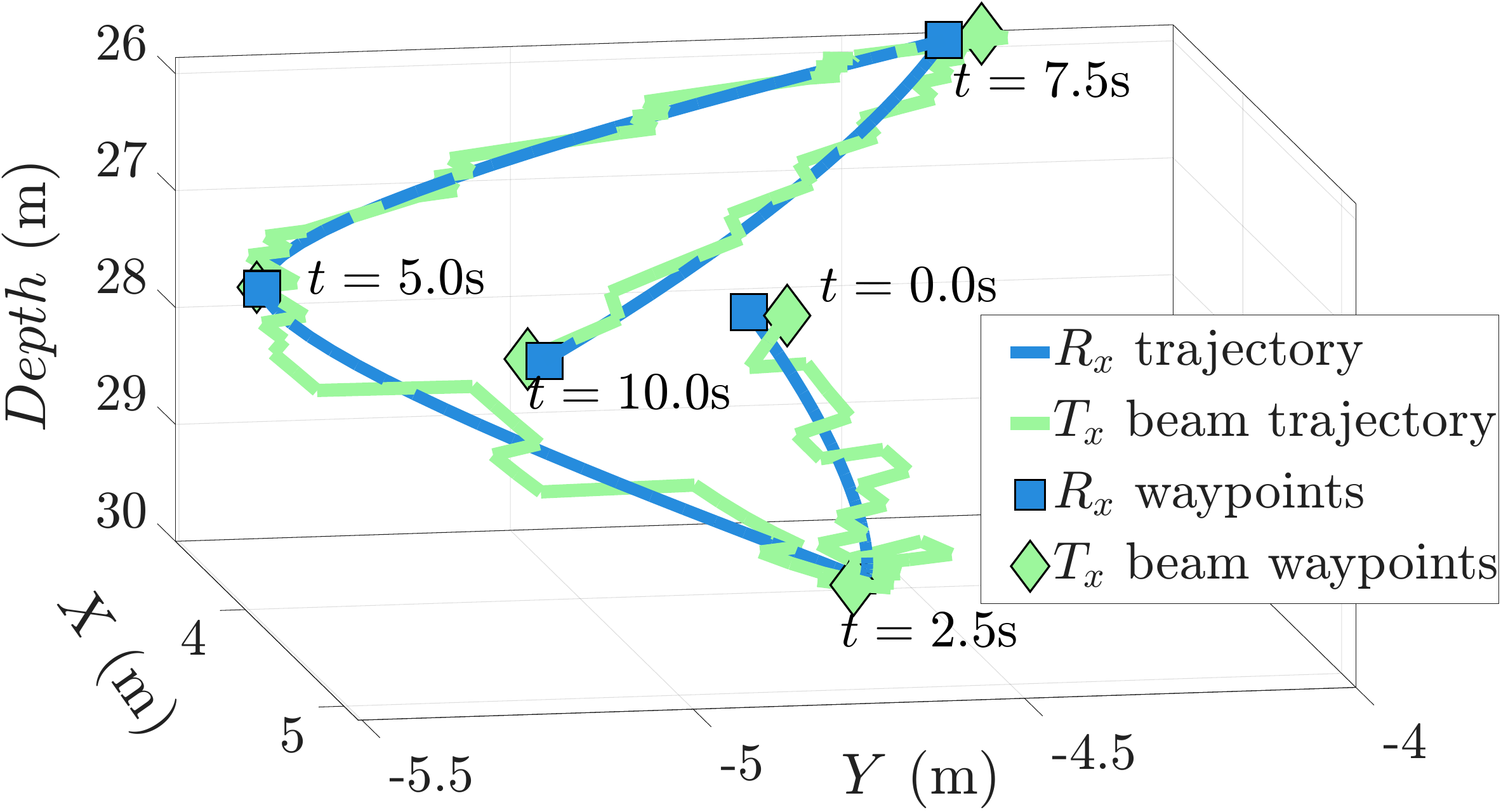}
\caption{First test episode under current-free conditions, showing the trajectory of the R$_x$-AUV through the planned waypoints (blue) alongside the tracking path of the T$x$-AUV (green).} 
\label{fig:testcase1_beamtracking}
\vspace{-0.35cm}
\end{figure}
\begin{figure}[!t]
    \centering
    \subfloat[]{
        \includegraphics[width=1\columnwidth]{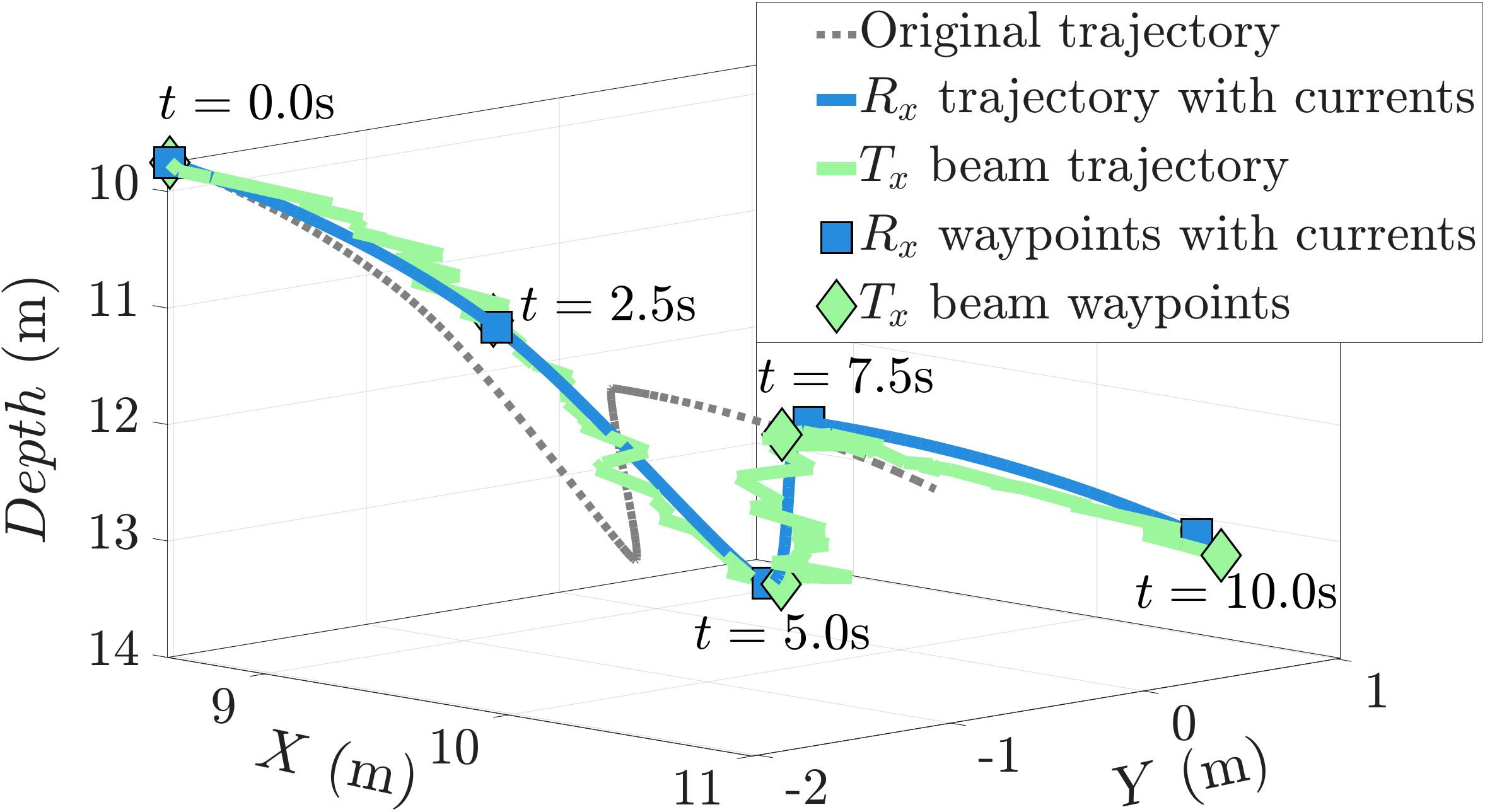}
        \label{fig:test_case2_beam_tracking}
    }
    
    \vspace{0.7cm} 
    
    \subfloat[]{
        \includegraphics[width=1\columnwidth]{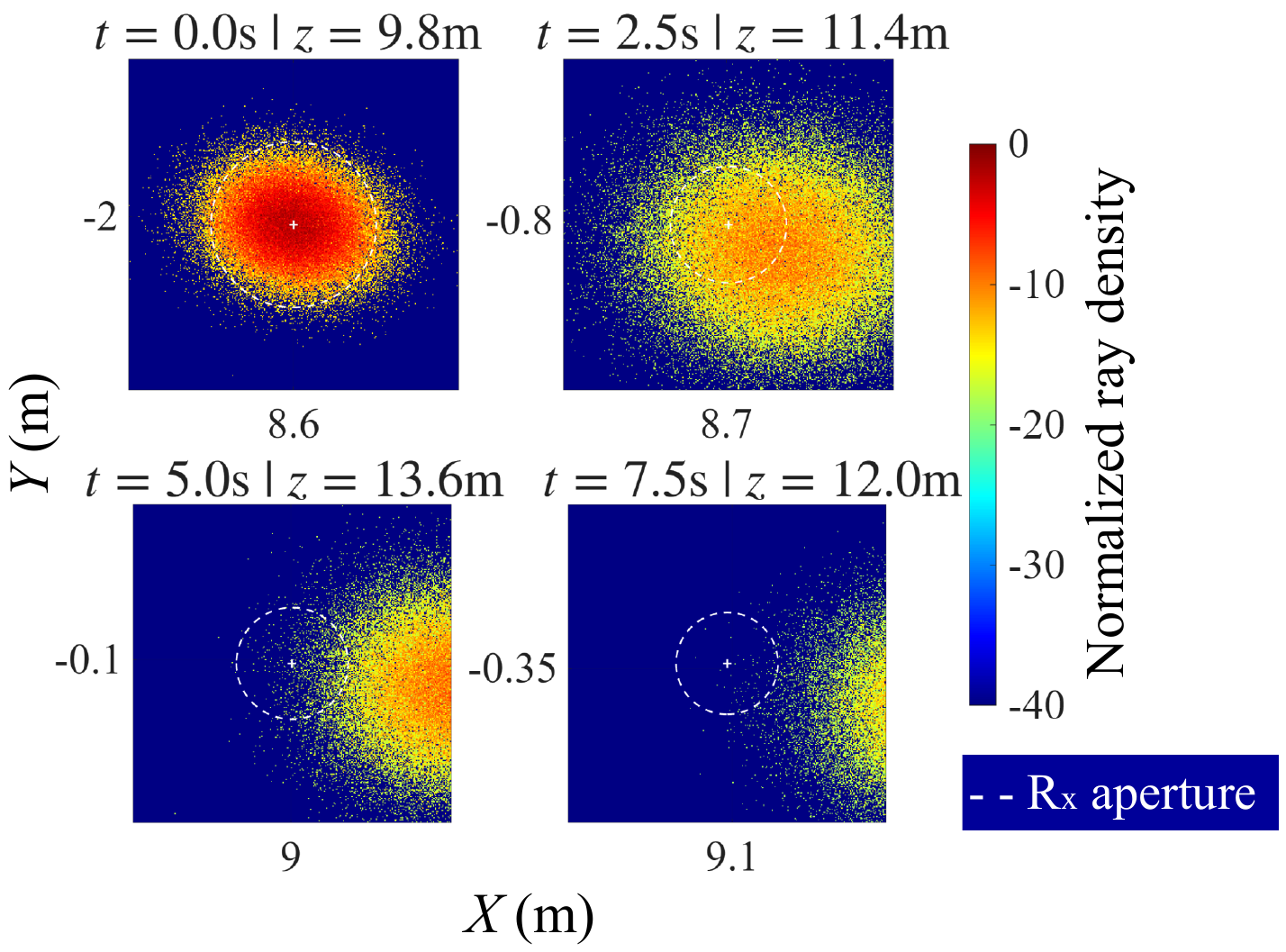}
        \label{fig:test_case2_waypoints}
    }
    
    \caption{Second test episode under induced-current conditions. (a) Actual R$_x$-AUV trajectory (blue) deviating from the planned trajectory (dotted black), alongside the T$_x$ beam path (green).  (b) Evolution of beam alignment at different time steps, where the agent increases beam divergence to maintain connectivity.}
    \vspace{-0.35cm}
    \label{fig:second_test_case}
\end{figure}

Following training, we evaluate the system on unseen trajectories. Fig. \ref{fig:testcase1_beamtracking} illustrates a first test episode under positional certainty conditions, where the R$_x$-AUV trajectory is closely followed by the agent's beam-steered path, despite minor path oscillations caused by the discrete step decision-making process. A second test episode assesses the system's resilience to ocean currents. 
As shown in Fig.~\ref{fig:test_case2_beam_tracking}, the R$_x$-AUV path deviates from its planned path due current-induced drift. Nevertheless, the agent actively tracks the drifting target, demonstrating the successful integration of environmental data. Because this continuous drift degrades alignment efficiency and causes the beam to progressively deviate from the receiver aperture, the agent autonomously increases the beam divergence to maintain connectivity. Fig.~\ref{fig:test_case2_waypoints} illustrates this adaptive response, which yields a more diffuse beam over time. 

\begin{figure}[!t]
\centering
\includegraphics[width=1\columnwidth]{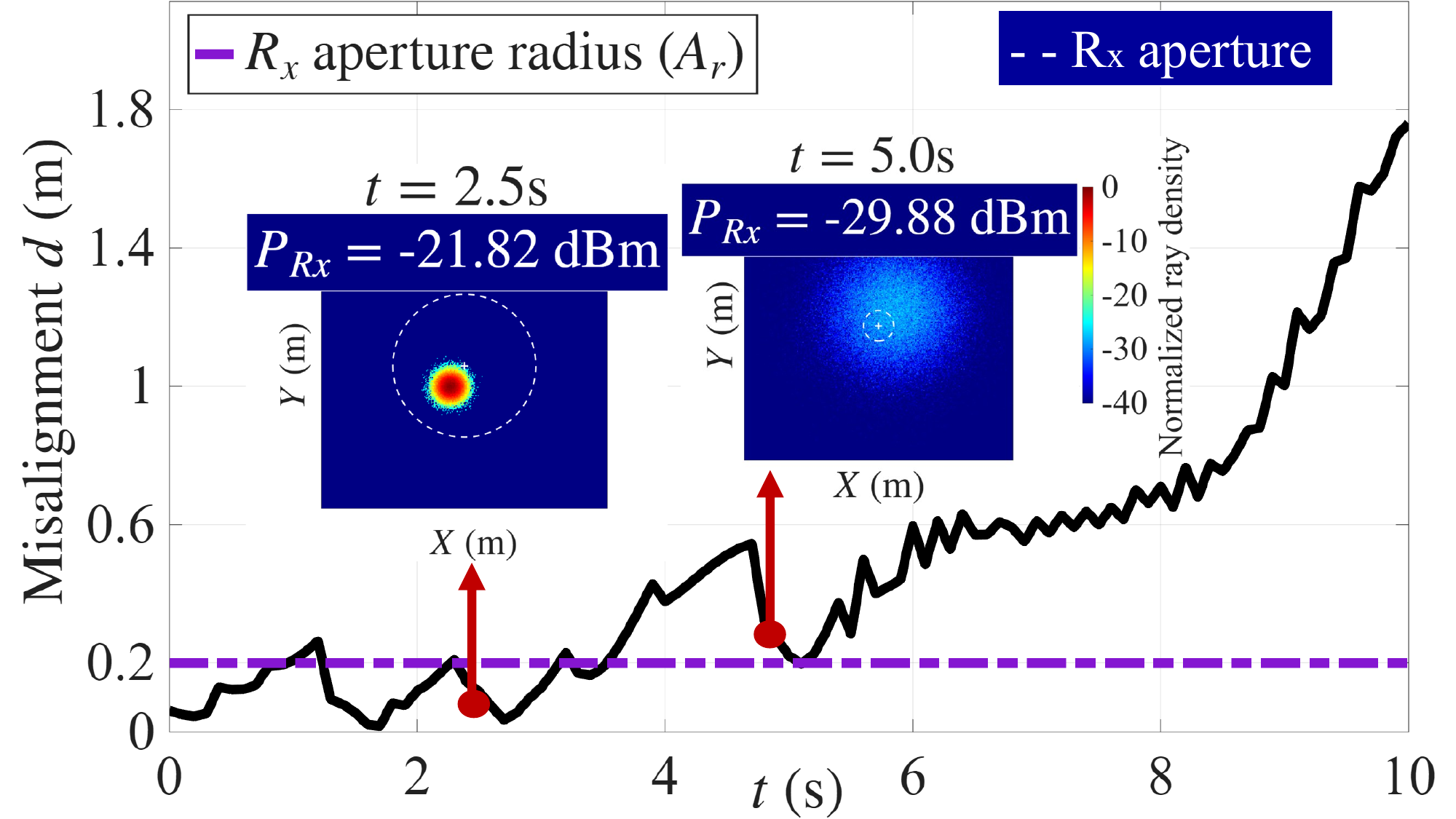}
\hspace{0.1cm}
\caption{Third test episode illustrating the positional misalignment relative to $A_r$, demonstrating that the agent narrows the beam when alignment is precise.}
\label{fig:test_case3_waypoints}
\end{figure}

To further analyze the agent's dynamic beam management capabilities under drift conditions, we evaluate a third test episode, depicted in Fig. \ref{fig:test_case3_waypoints}. At 2.5~s, the positional misalignment remains tightly within the overlap region defined in Fig.~\ref{fig:reward_condition}, specifically below $A_r$. Leveraging this precise alignment, the agent narrows the beam to maximize the received power, achieving -21.82~dBm. Beyond 3.5~s, tracking stability deteriorates and the agent responds through increasing beam divergence to preserve the communication link.

\begin{figure}[!t]
\centering
\includegraphics[width=1\columnwidth]{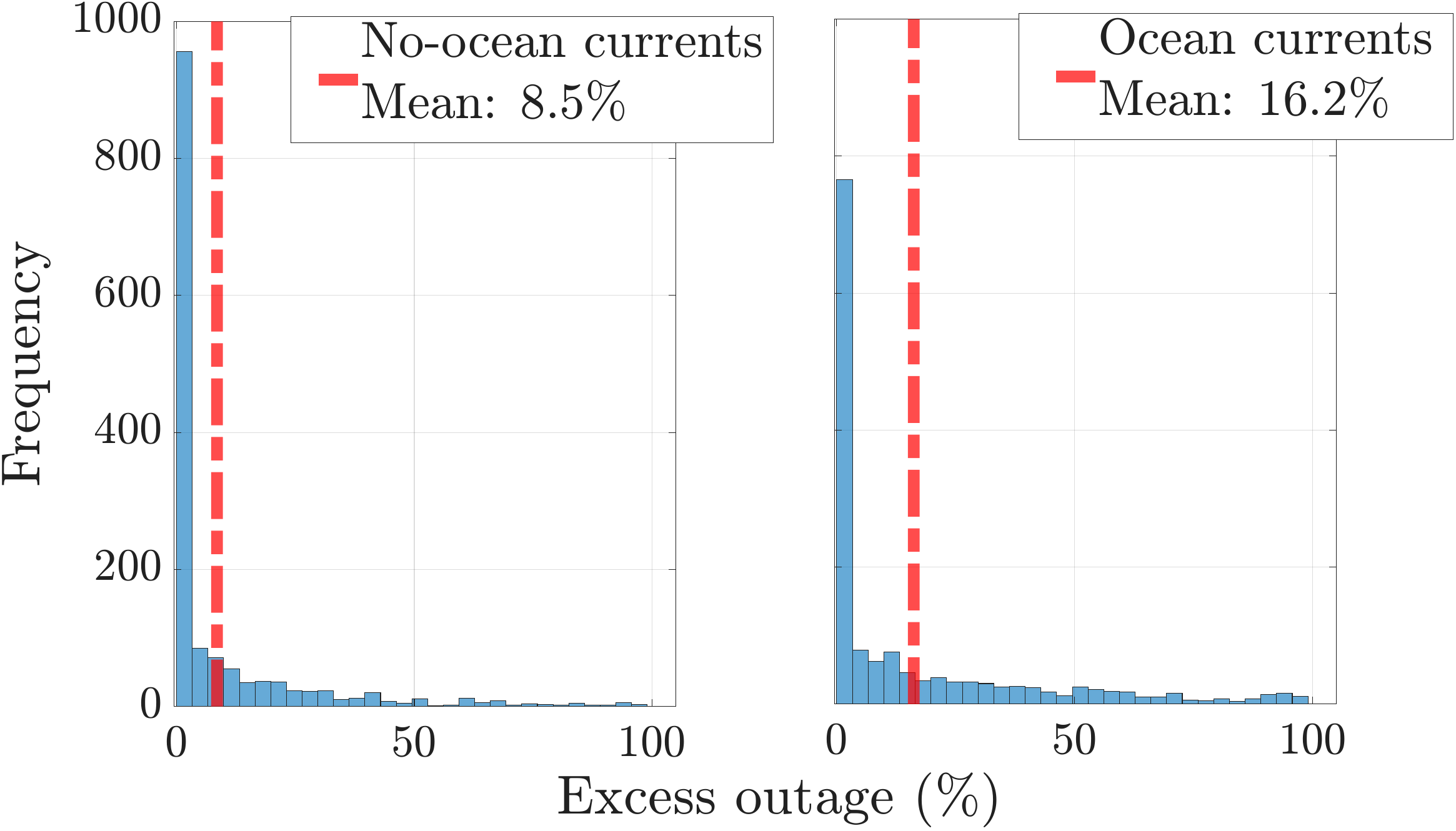}
\caption{Assessment of the DRL agent's tracking accuracy across 1,500 episodes for both operational scenarios.}
\label{fig:resutls_global}
\end{figure}

Moving beyond individual test cases, the agent’s overall performance is assessed across 1,500 unseen episodes by measuring the percentage of simulation time experiencing communication outages ($SNR~<\text{10 dB}$). To isolate outages caused by depth-dependent channel effects, which are beyond the agent’s control, we compare the agent's per-episode outage duration against a baseline that assumes perfect R$_x$-AUV tracking. The difference between both is the excess outage metric, which quantifies the fraction of simulation time where connectivity is lost exclusively due to pointing errors. As depicted in Fig. \ref{fig:resutls_global}, the histograms for both scenarios are positively skewed with the primary peak located at the lower bound, corresponding to episodes with zero tracking-induced signal loss. Under current-free conditions, the mean excess outage is 8.5\%, indicating that, on average, pointing errors account for 0.8~s of signal loss during the total 10-second simulation. However, in current-induced trajectories, this metric increases to 16.2\%. This performance degradation reflects the increased complexity of the task, as the agent must simultaneously manage pointing control, estimate current-induced drift, and predict the R$_x$-AUV's actual position.

\section{Conclusion}
\label{sec:conclusion}
This work presented a 3D DRL framework to address transmitter-receiver misalignment between mobile AUVs. To model this environment, we developed a 3D UWOC channel incorporating depth-dependent impairments, and a discrete-ray mechanism to simulate pointing loss. Evaluating 1,500 trajectories, we assessed the framework using the excess outage metric, which strictly accounts for the pointing accuracy of the agent. Under positional certainty, the agent successfully tracked the R$_x$, limiting the mean excess outage to 8.5\% when compared to ideal conditions without pointing errors. In the scenario where the R$_x$'s trajectory is perturbed by ocean currents, the system constrained this metric to 16.2\%. These results demonstrate that the agent learned a baseline policy to estimate oceanic drift and preserve link reliability by simultaneously controlling beam direction and divergence.

\section*{Acknowledgment}
This research is supported by the AVALON project under European Union's European Defence Fund under Grant Agreement No. 101168152 and by the Spanish Ministry of Science, Innovation and Universities MICIU/AEI/10.13039/501100011033 through the Ram\'{o}n y Cajal grant RYC2024-051003-1.

\bibliographystyle{IEEEtran}
\bibliography{IEEEabrv,bib}

\begin{thebibliography}{10}
\providecommand{\url}[1]{#1}
\csname url@samestyle\endcsname
\providecommand{\newblock}{\relax}
\providecommand{\bibinfo}[2]{#2}
\providecommand{\BIBentrySTDinterwordspacing}{\spaceskip=0pt\relax}
\providecommand{\BIBentryALTinterwordstretchfactor}{4}
\providecommand{\BIBentryALTinterwordspacing}{\spaceskip=\fontdimen2\font plus
\BIBentryALTinterwordstretchfactor\fontdimen3\font minus \fontdimen4\font\relax}
\providecommand{\BIBforeignlanguage}[2]{{%
\expandafter\ifx\csname l@#1\endcsname\relax
\typeout{** WARNING: IEEEtran.bst: No hyphenation pattern has been}%
\typeout{** loaded for the language `#1'. Using the pattern for}%
\typeout{** the default language instead.}%
\else
\language=\csname l@#1\endcsname
\fi
#2}}
\providecommand{\BIBdecl}{\relax}
\BIBdecl

\bibitem{Gorodilov2025}
A.~Gorodilov, I.~Zhilin, A.~Sakhnini, J.~Simonjan, and I.~Akyildiz, ``Unisdm proof of concept: A multimode software defined modem for underwater communications,'' in \emph{2025 IEEE Wireless Communications and Networking Conference (WCNC)}, 2025, pp. 1--3.

\bibitem{saeed2019underwater}
N.~Saeed, A.~Celik, T.~Y. Al-Naffouri, and M.-S. Alouini, ``Underwater optical wireless communications, networking, and localization: A survey,'' \emph{Ad hoc networks}, vol.~94, p. 101935, 2019.

\bibitem{vali2017use}
Z.~Vali, A.~Gholami, D.~G. Michelson, Z.~Ghassemlooy, M.~Omoomi, and H.~Noori, ``Use of gaussian beam divergence to compensate for misalignment of underwater wireless optical communication links,'' \emph{IET Optoelectronics}, vol.~11, no.~5, pp. 171--175, 2017.

\bibitem{shakeera2023ml}
S.~Shakeera, V.~B.~N. Jyothi, and H.~Venkataraman, ``{ML}-based techniques for prediction of ocean currents for underwater vehicles,'' in \emph{2023 11th International Symposium on Electronic Systems Devices and Computing (ESDC)}, vol.~1.\hskip 1em plus 0.5em minus 0.4em\relax IEEE, 2023, pp. 1--6.

\bibitem{romdhane2022reinforcement}
I.~Romdhane and G.~Kaddoum, ``A reinforcement-learning-based beam adaptation for underwater optical wireless communications,'' \emph{IEEE Internet of Things Journal}, vol.~9, no.~20, pp. 20\,270--20\,281, 2022.

\bibitem{ishida2025adaptive}
T.~Ishida, C.~B. Naila, and H.~Okada, ``Adaptive beam steering and divergence control for underwater optical wireless communication using reinforcement learning,'' \emph{IEEE Photonics Journal}, 2025.

\bibitem{shin2024multidimensional}
H.~Shin, S.~Baek, and Y.~Song, ``Multidimensional beam optimization in underwater optical wireless communication based on deep reinforcement learning,'' \emph{IEEE Internet of Things Journal}, vol.~11, no.~17, pp. 28\,623--28\,634, 2024.

\bibitem{nrl_gofs31}
\BIBentryALTinterwordspacing
{Naval Research Laboratory}, ``Global ocean forecast system (gofs) 3.1,'' Dataset, 2014--2021. [Online]. Available: \url{https://www.hycom.org/dataserver/gofs-3pt1/analysis}
\BIBentrySTDinterwordspacing

\bibitem{cox2014simulating}
W.~Cox and J.~Muth, ``Simulating channel losses in an underwater optical communication system,'' \emph{Journal of the Optical Society of America A}, vol.~31, no.~5, pp. 920--934, 2014.

\bibitem{koutsonas2026performance}
E.~Koutsonas, A.~Pitilakis, O.~Tsilipakos, T.~A. Tsiftsis, and A.~A.~A. Boulogeorgos, ``Performance analysis of underwater optical wireless communications under the joint impact of disorientation and misalignment,'' in \emph{Proceedings of the 20th European Conference on Antennas and Propagation (EuCAP 2026)}, March 2026, accepted for publication.

\bibitem{jerlov1976marine}
N.~G. Jerlov, \emph{Marine optics}.\hskip 1em plus 0.5em minus 0.4em\relax Elsevier, 1976, vol.~14.

\bibitem{Haltrin}
V.~Haltrin, ``Chlorophyll-based model of seawater optical properties,'' \emph{Journal of Applied Optics}, vol.~38, no.~33, pp. 6826--6832, 1999.

\bibitem{Deng}
R.~Deng, Y.~He, Y.~Quin, Q.~Cheng, and L.~Chen, ``Pure water absorption coefficient measurement after eliminating the impact of suspended substance in spectrum from 400 nm to 900 nm,'' \emph{Journal of Remote Sensing}, pp. 1007--4619, 2012.

\bibitem{johnson2015optical}
L.~J. Johnson, ``Optical property variability in the underwater optical wireless channel,'' Ph.D. dissertation, University of Warwick, 2015.

\bibitem{Utiz}
J.~Utiz, H.~Clustré, A.~Morel, and S.~Hoooker, ``Vertical distribution of phytoplankton communities in open ocean: An assessment based on surface chlorophyll,'' \emph{Journal of Geophysical Research}, vol. 111, pp. 1--23, 2006.

\bibitem{elamassie2020vertical}
M.~Elamassie and M.~Uysal, ``Vertical underwater visible light communication links: Channel modeling and performance analysis,'' \emph{IEEE Transactions on Wireless Communications}, vol.~19, no.~10, pp. 6948--6959, 2020.

\bibitem{TEOS10_Software}
\BIBentryALTinterwordspacing
{IOC}, {SCOR}, and {IAPSO}. (2011) {TEOS}-10 {Software}: {Gibbs} {Seawater} ({GSW}) {Oceanographic} {Toolbox}. Intergovernmental Oceanographic Commission, Scientific Committee on Oceanic Research, and International Association for the Physical Sciences of the Oceans. [Online]. Available: \url{https://www.teos-10.org/software.htm}
\BIBentrySTDinterwordspacing

\bibitem{tayebnaimi2024scintillation}
S.~Tayebnaimi and K.~Kiasaleh, ``Scintillation index analysis for multi-wavelength gaussian beams in turbulent underwater channels,'' \emph{Optical Engineering}, vol.~63, no.~12, pp. 128\,105--128\,105, 2024.

\bibitem{zayed2025modeling}
M.~M. Zayed and M.~Shokair, ``Modeling and simulation of optical wireless communication channels in iout considering water types turbulence and transmitter selection,'' \emph{Scientific Reports}, vol.~15, no.~1, 2025.

\bibitem{mahmutouglu2020investigation}
Y.~Mahmuto{\u{g}}lu, C.~Albayrak, and K.~T{\"u}rk, ``Investigation of underwater wireless optical communication channel capacity for different environment and system parameters,'' \emph{Hittite Journal of Science and Engineering}, vol.~7, no.~4, pp. 279--285, 2020.

\end{thebibliography}
\end{document}